# Deep learning scheme for recovery of broadband microwave photonic receiving systems in transceivers without expert knowledge and system priors

SHAOFU XU,[1] RUI WANG,[1] JIANPING CHEN,[1] LEI YU,[1] AND WEIWEN ZOU[1,*]

[1] *State Key Laboratory of Advanced Optical Communication Systems and Networks, Intelligent Microwave Lightwave Integration Innovation Center (iMLic), Department of Electronic Engineering, Shanghai Jiao Tong University, 800 Dongchuan Road, Shanghai 200240, China.*
*\*wzou@sjtu.edu.cn*

**Abstract:** In regular microwave photonic (MWP) receiving systems, broadband signals are processed in the analog domain before they are transformed to the digital domain for further processing and storage. However, the quality of the signals may be degraded by defective photonic analog links, especially in a complicated MWP system. Here, we show a unified deep learning scheme that recovers the distorted broadband signals as they are transformed to the digital domain. The neural network could automatically learn the end-to-end inverse responses of the distortion effects of actual photonic analog links from data without expert knowledge and system priors. Hence, by shifting or augmenting the datasets, the neural network is potential to be generalized to various MWP receiving systems. We conduct experiments by nontrivial MWP systems with complicated waveforms. Results validate the effectiveness, general applicability and the noise-robustness of the proposed scheme, showing its superior performance in practical MWP systems. Therefore, the proposed deep learning scheme facilitates the low-cost performance improvement of MWP receiving systems, as well as the next-generation broadband transceivers, including radars, communications, and microwave imaging.

## 1. Introduction

Microwave photonics (MWP) is a powerful candidate for the next-generation information systems because of its capability to process ultra-wideband signals [1, 2]. However, a fully-functional MWP systems are usually constructed by complicated analog links and lots of photonic components. Due to the imperfect properties of analog links and photonic components, signals can be distorted and the system fidelity is inevitably degraded [3, 4]. To solve the problem of signal distortion, traditional methods refine the analog links by analytically modeling and compensating the distortion effects of the defective photonic components [5-7]. However, this strategy requires considerable expert knowledge and auxiliary assistant systems, which involve high deployment costs. Moreover, there is always a disparity among the photonic components, posing additional challenges in terms of the efficiency and robustness of the traditional strategy.

Data-driven autonomous feature learning is one of the main factors contributing toward the adoption of deep learning across disciplines and applications [8]. With appropriate architectures, hyperparameters, and training methods, neural networks can learn the mapping between the input data and the output automatically; thus, no additional expert knowledge is required. By following this strategy, deep learning has achieved outstanding performance in diverse tasks across many data categories, including sequences [9, 10], waveforms [11], images [12-14], videos [15], and other irregular formats [16-18]. Particularly in the case of regression tasks, deep neural networks are used to effectively conduct image reconstruction [19], super-resolution [20], denoising, and de-mosaicking [21], verifying the universal function approximation ability of neural networks [22]. Most recently, we have adopted deep learning to enhance the performance of photonic analog-to-digital converter: the capability and high accuracy of neural networks are experimentally demonstrated with sine-alike waveforms [23]. Therefore, it is potential to apply deep neural networks to recover the distortions of complicated broadband signals in general MWP processing systems.

However, there are some hindrances to directly adopt conventional deep learning strategy to MWP systems. In the training phase of the conventional strategy, the supervising ground-truth data are prepared in datasets [24, 25] by fine acquisition systems such as high-performance cameras and microscopes. The degraded data are generated from the ground-truth data through theoretical transform models or by adding modeled noise, and the neural networks are trained to recover the degraded data as finely as possible. This strategy is hindered in the case of broadband signal recovery because (1) accurate modeling of all the components in an analog link as well as their distortion effects is challenging; (2) the analog systems can deviate from the theoretical model, and the theoretically trained neural networks may perform poorly in practical systems; and (3) the acquisition of the ground-truth data depends on a fine acquisition system that is rarely available for broadband signal applications.

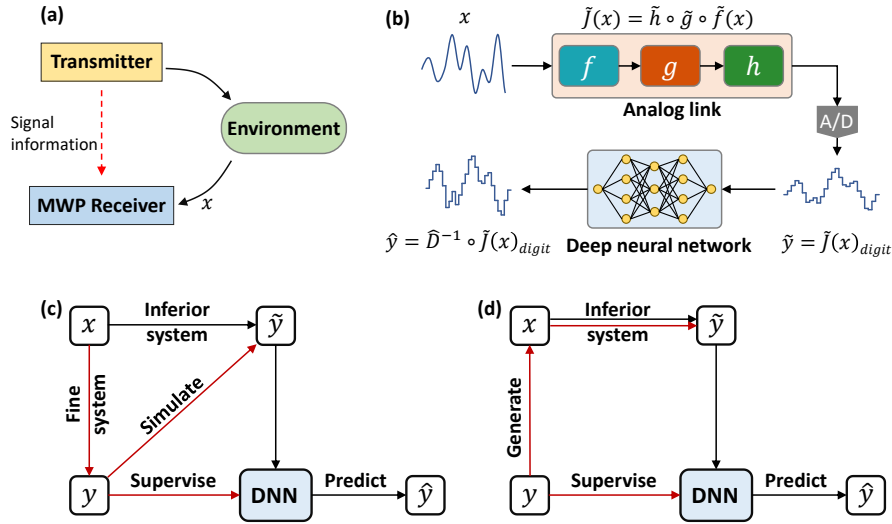

Fig. 1. Schematics of deep learning scheme for broadband signal recovery. (a) Schematic of a typical MWP transceiver system. (b) Schematic of the signal flow in the MWP receiver. The original analog signal $x$ passes through the defective photonic analog link (three stages of responses for example) and is transformed to the digital domain by an analog-to-digital converter (A/D). Then, a DNN is deployed to recover the distorted digital signal $\tilde{y}$. After the DNN-learned inverse response is generated, the digital signal is recovered as $\hat{y}$. In the figure, $\tilde{h} \circ \tilde{g} \circ \tilde{f}(\cdot)$ is the defective version of $h \circ g \circ f(\cdot)$ and $\widehat{D}^{-1}(\cdot)$ represents the DNN-learned end-to-end inverse response. (b)(c) Comparison of strategies adopted in conventional deep learning data recovery and the proposed scheme. The two strategies have the same inference phase (depicted with black lines): the original analog signal $x$ is processed by an inferior system to obtain the degraded data $\tilde{y}$. The trained DNN predicts the good data $\hat{y}$ from the degraded data. However, they have different training phases. The red lines represent the training prerequisites and $y$ is the supervising ground truth in both strategies.

This study proposes and demonstrates a unified deep learning scheme to recover the broadband signals from defective MWP receiving systems. Instead of analytically modeling each component in analog links, the proposed scheme uses the distorted signals and supervising signals to train a neural network. Therefore, the neural networks can learn the end-to-end inverse responses of the MWP systems and recover the distorted signals effectively. Moreover, this scheme does not require fine acquisition systems. Therefore, it is suitable for MWP applications where fine acquisition systems are expensive and rare. Experiments also testify the generalizability and noise-robustness of the scheme, indicating the potential applications in more various MWP receiving systems under practical noisy conditions. By shifting or augmenting the training datasets, the neural network is able to evolve to adapt to various MWP receiving systems and applications.

## 2. Principles and concepts

The basic concept of the proposed scheme is shown in Fig. 1(a) and 1(b). In a typical MWP transceiver system, the transmitting signals are generated by cooperative transmitter. Therefore, it can provide signal information to construct the training datasets regarding different applications. Inside the MWP receiving system, complicated analog links are deployed to perform signal pre-processing in the analog domain before the signals are transformed to the digital domain for further processing, storage, and display. When the original analog signal $x$ enters an analog link that is defective, it suffers from the distortion effects given by $\tilde{J}(x) = \tilde{h} \circ \tilde{g} \circ \tilde{f}(x)$ (circles here represent function compositions). After transformation to the digital domain, the distorted signal can be recovered by the deep neural network (DNN). The DNN is trained to learn the end-to-end inverse response of the distortion effects of the analog link $\hat{D}^{-1}$, and it yields the recovered signal in the digital domain, $\hat{y} = \hat{D}^{-1} \circ \tilde{J}(x)_{digit}$.

We show the strategy difference of conventional deep learning data recovery and the proposed scheme in Fig. 1(c) and 1(d), the signal flows in the inference phases of the conventional strategy and the proposed scheme are the same; both use DNNs to predict the good data from the degraded data acquired by the inferior systems. However, they are different in terms of the training prerequisites. In the conventional strategy, fine acquisition systems are necessary because the supervising ground-truth data can be only acquired by the fine acquisition systems. The degraded data are simulated by imposing theoretical inverse transformations or modeled noise on the ground truth. However, the traditional strategy is strongly hindered in the case of MWP systems. Therefore, the original analog signal $x$ is artificially generated from the ground-truth data. When the analog signal passes through the practical inferior system, the distortions in the degraded data $\tilde{y}$ reflect the actual distortion effects of the analog link. Consequently, through the supervision of the ground-truth data $y$, the neural network can learn the actual end-to-end inverse response of the analog link and recover these signals effectively. Compared with applications where signals are represented with high-dimensional structured data such as images and videos, it is convenient to artificially generate one-dimensional broadband signals with current waveform generation technologies. Thus, artificially generated signals can be employed to determine the actual distortion effects of various MWP systems.

**Table. 1 Network structure and hyperparameters of RAE.**

| Layer name | Conv. window | Channel no. | Striding | Padding | Activation |
|---|---|---|---|---|---|
| Input | 1×5 | 1 to 32 | 2 | "SAME" | ReLU |
| 1st conv. | 1×5 | 32 to 34 | 2 | "SAME" | ReLU |
| 2nd conv. | 1×3 | 34 to 38 | 2 | "SAME" | ReLU |
| 3rd conv. | 1×3 | 38 to 44 | 2 | "SAME" | ReLU |
| 4th conv. | 1×3 | 44 to 44 | 1 | "SAME" | ReLU |
| 1st de-conv. | 1×5 | 44 to 38 | 2 | "SAME" | Tanh |
| 2nd de-conv. | 1×5 | 38 to 34 | 2 | "SAME" | Tanh |
| 3rd de-conv. | 1×5 | 34 to 32 | 2 | "SAME" | Tanh |
| 4th de-conv | 1×7 | 32 to 1 | 2 | "SAME" | Tanh |

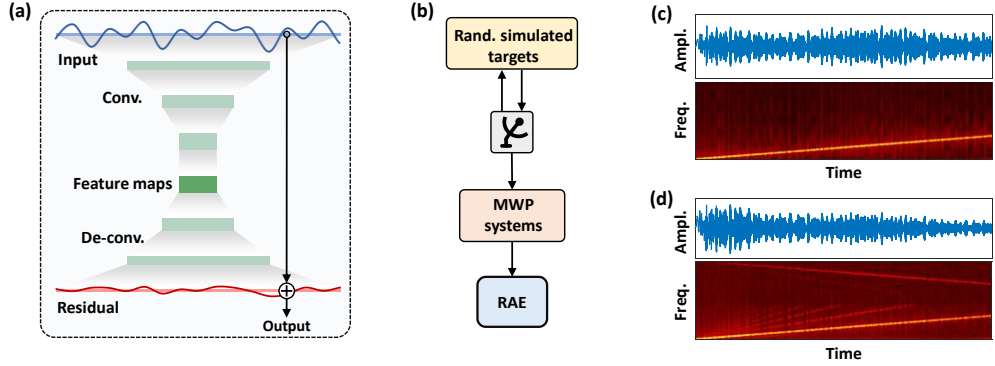

Fig. 2. Experimental implementations. (a) Model of the RAE. The input and output data of RAE are time-domain waveforms. With several convolutional layers (Conv.), the input data are transformed to feature maps. With several de-convolutional layers (De-conv.) [28], these feature maps are reconstructed into residual data. By adding the residual data to the original input data, the output is yielded. (b) Schematic of the experimental setup. We Monte-Carlo simulate complicated broadband radar echo waveforms. The transmitting signals are frequency-modulated (LFM, Costas) signals, and they are bounced back by randomly (rand.) simulated targets to form complicated waveform datasets. The simulated waveforms are generated in the analog domain via an AWG and passed through the MWP systems (PPS, PADC). Then, the RAE is applied to recover the distorted data. (c)(d) Time-domain plots and time-frequency domain plots of a radar echo before and after the MWP systems, respectively.

## 3. Experimental implementations

The neural network employed in this study is constructed according to the concept of all-convolutional autoencoders [26] together with the residual neural networks [27], as shown in Fig. 2(a). The adopted neural network is denoted as the residual autoencoder (RAE). All the synaptic links in the RAE are convolutional for adaptation to the varying input lengths. The input and output of the RAE are the time-domain waveforms, and the RAE learns the end-to-end inverse response of the distortion effects of analog links automatically. The network structure and hyper-parameters of RAE are list in Table. 1. When the RAE is supposed to be trained to adapt to a new dataset, the network structure and hyperparameters are unchanged. Note that the RAE has a shrinking-and-expanding structure, where several convolutional layers transform the input time-domain waveform to feature maps. These feature maps are then de-convoluted to reconstruct the time-domain output waveform. The RAE is implemented with TensorFlow framework in Python on an Intel i5-8400 computer with a deep learning accelerator (dual Nvidia GTX-1080-ti GPUs).

To investigate the general feasibility of the RAE for the broadband signal recovery of various receiving systems, we set up experiments with two nontrivial MWP systems. One of them is the photonic parametric sampling (PPS) system [29] and the other is the photonic-assisted analog-to-digital conversion (PADC) system [30]. PADC system is used in this study as an example because of its special and intricate distortion effects. Compared with [23], the application scenario of PADC here is generalized to complicated broadband signals. In Fig. 3, the experimental setup of MWP systems are illustrated and all details of microwave photonic components are explained. The PPS and PADC systems differ in terms of their principles and distortion effects; hence, they are appropriate for demonstrating the general adaptiveness of the RAE.

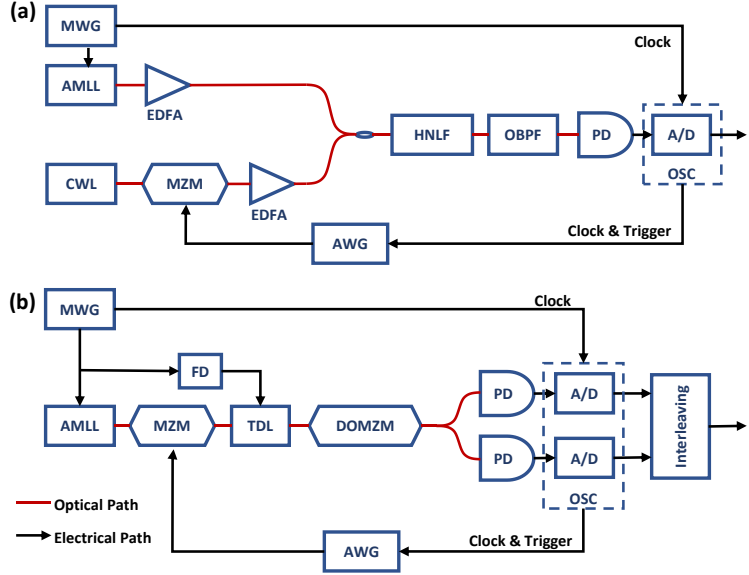

Fig. 3. Experimental setups of MWP systems. (a) PPS system and (b) PADC system. MWG, microwave generator, Keysight E8257D; AMLL, actively mode-locked laser, CALMAR PSL-10-TT; CWL, continuous-wave laser, Alnair Labs TLG-200; EDFA, erbium-doped fiber amplifier, Calmar AMP-ST30 and Calmar PCS-2; MZM, Mach-Zehnder modulator, Optilab IMC-1550; HNLF, high-nonlinearity fiber, TZ01563DA02; OBPF, optical band-pass filter, Alnair Labs CVF-220CL; PD, photon-detector, Discovery DSC50S; A/D, analog-to-digital conversion; OSC, oscilloscope, Keysight DSO-S 804A; AWG, arbitrary waveform generator, Keysight M8195A.. DOMZM, Dual-output MZM, PHOTLINE AX-1x2-0MsSS-20-SFU-LV; FD, frequency divider; TDL, tunable delay line, General Photonics VDL-001.

As shown in Fig. 2(b), the complicated waveforms of the broadband radar echoes are chosen to be the input signals of the MWP systems. The radar echoes can be highly complicated in broadband situations because the targets become intricate under broadband signal illuminations [31, 32]. Therefore, it is appropriate to check the adaptiveness of the RAE to the practical complicated waveforms. To simulate and generate the radar echo waveforms, the transmitting waveforms are chosen from two categories: linearly frequency-modulated (LFM) and frequency-jumping (Costas). The parameters of targets are generated randomly: number of scattering centers (from 1 to 12), reflective cross section of each scattering center (from 0 to 1, normalized), spatial location of each scattering center (within 10-m range), velocity of the reflective targets (from 0 to 300 m/s), and rotation speed (from 0 to 0.2 rad/s). Based on the multi-scattering-center theory of broadband radar, the echo waveforms can be generated with chosen transmitting waveforms and target parameters. Figure 2(c) and 2(d) show an example of generated waveforms of simulated radar echoes before and after the distortion effects of the MWP analog links, respectively. From the time-domain and time-frequency-domain plots, the defective analog link nonlinearly distorts the amplitude of the signal and introduces frequency pseudo-image/harmonics.

## 4. Experimental results

Based on different waveform categories and different MWP systems, we acquire individual datasets to train the RAE. Note that the noise floor of the AWG-generated signals is relatively heavy. In order to build a clean training set, we utilized 128-time acquisitions to mitigate the random noise. Figure 4 shows the effect of averaging. The data is acquired by the oscilloscope directly. Without averaging, the power density of noise is 28 dB below signal. With averaging, the noise floor is decreased to 37 dB below the signal. Note that the averaging does not influence the distortion effects of the signals: therefore, we could acquire datasets with less noise but the distortion effects are unchanged. The clean datasets can allow us to focus on the RAE effectiveness of distortion recovery rather than the influence of random noise. Using this method, we acquire three datasets: LFM echoes before and after PSS system, Costas echoes before and after PSS system, LFM echoes before and after PADC system. Each dataset contains 250 examples, which are divided to training dataset (200 examples) and validating dataset (50 examples).

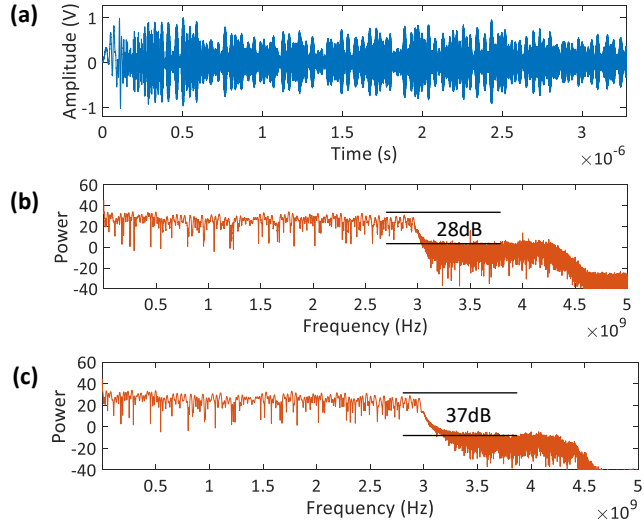

Fig. 4 Random noise mitigation of AWG generated signals. (a) Time domain of the AWG generated LFM signal. (b) Frequency domain of non-average LFM signal. (c) Frequency domain of 128-times averaged LFM signal.

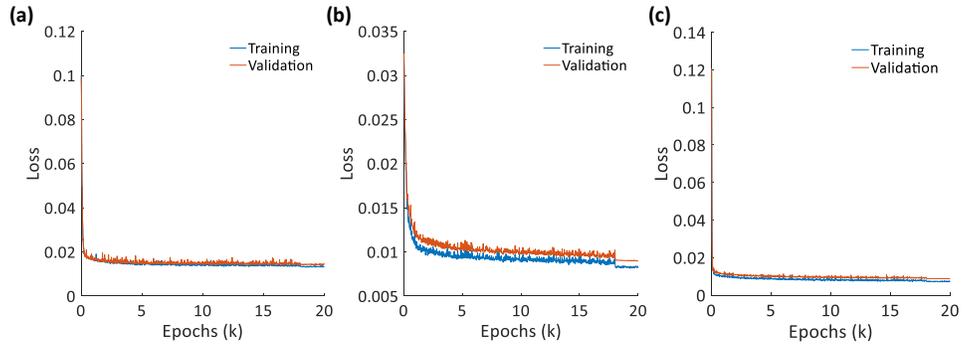

Fig. 5. Loss curves during the RAE training based on different datasets. (a) LFM-PPS dataset. (b) Costas-PPS dataset. (c) LFM-PADC dataset. The loss curves show that the validation losses drop consistently with the training losses, and they verify that the RAE fits the three datasets properly.

By only altering the training dataset, the RAE is trained to adapted to different waveforms and various analog links. The loss function of training is defined as the absolute error between network output and reference data:

$$Loss(\Theta) = \frac{1}{L}\sum_{l=1}^{L} | Y_l^{\Theta} - Y_l^{ref} |, \tag{1}$$

where $Loss(\Theta)$ is the loss function with network parameters of $\Theta$. $L$ is the length of output data. $Y^{\Theta}$ and $Y^{ref}$ are the network output data and the reference data, respectively. The training is conducted by iterating "Adam" algorithm [33] for 20k times. After 18k iterations, the learning rate is decayed from 1e-3 to 1e-4. Figure 5 depicts the training loss curve and validation loss curve during training. The two loss curves drop consistently, inferring that the RAE fits the three training datasets and validation datasets properly. In our experimental platform, the training procedure takes 29m56s. After training, the neural network can process input waveforms at a very high speed. We evaluated the neural network throughput on our experimental platform (dual GPUs), the neural network can output 86.16 Mega-points per second (Mpts/s). Suppose the memory depth of each waveform frame is 1 MB. The frame rate of the neural network can reach 336.6 frames per second (fps). Additionally, professional deep learning accelerators can further boost the throughput of the neural network [23]. Therefore, the appending of a neural network does not introduce heavy delay to the microwave photonic systems.

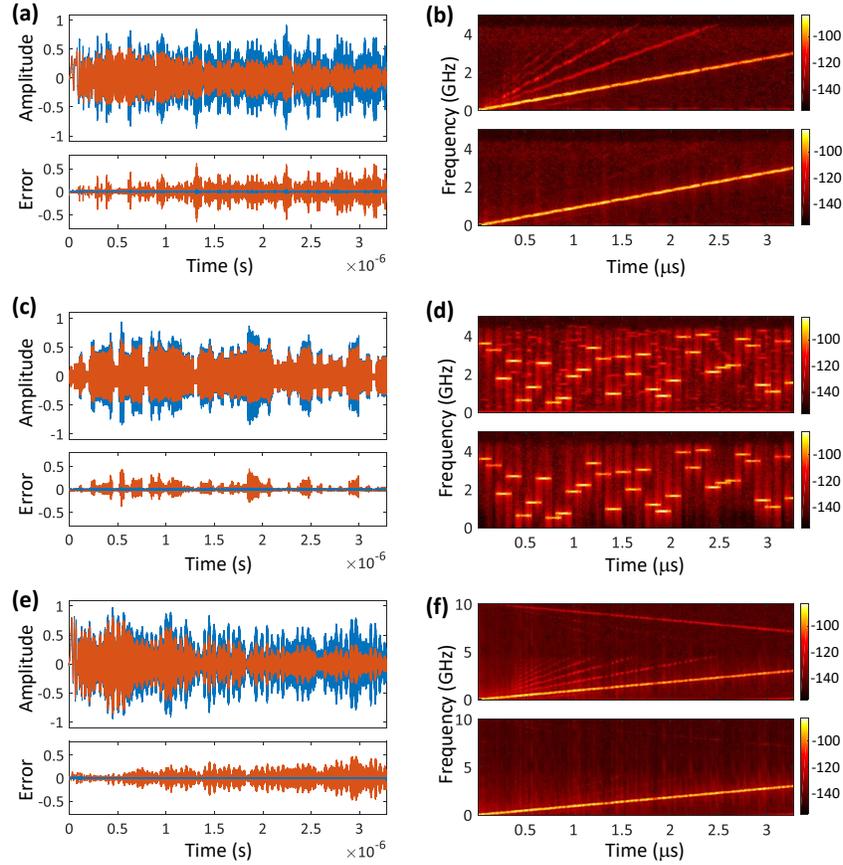

Fig. 6. Experimental results of broadband signal recovery. For every example, time-domain plots (a)(c)(e) and time-frequency-domain plots (b)(d)(f) are provided for better visibility of the recovery effects. In the time-domain plots, we give the waveforms of recovered/unrecovered signals in the upper subplots and the errors of recovered/unrecovered signals in the lower subplots. The orange and blue curves represent the results before and after recovery, respectively. In the time-frequency-domain plots, the unrecovered signals are shown in the upper subplots and the recovered are shown in the lower subplots. (a)(b) LFM echo waveforms with 3-GHz instantaneous bandwidth, and the distortion effects are from the PPS system. (c)(d) Costas echo waveforms with 3.5-GHz instantaneous bandwidth, distorted by the PPS system. (e)(f) LFM echo waveforms with 3-GHz instantaneous bandwidth, distorted by the PADC system.

Figure 6 shows experimental results of broadband signal recovery based on the RAE in time-domain and time-frequency-domain plots. An example of the recovery of a broadband LFM echo after the PPS system is shown in Fig. 6(a) and 6(b). Owing to the nonlinear amplitude response and frequency response of the PPS analog link, the broadband signal is heavily distorted. From the time-domain plot, we can roughly see that the high-amplitude parts are suppressed and the high-frequency parts are decayed; thus, the error between the distorted signal and the ground truth is significantly large. From the time-frequency-domain plot, these distortion effects lead to a series of harmonics. After the trained RAE, the distorted signal is recovered effectively, and the error is reduced to a fairly small value. We evaluate the mean square error (MSE) before and after signal recovery to characterize the performance of the RAE. In the case where LFM echoes pass through the PPS analog link, the MSE is averagely decreased by 17.46 dB. Figure 6(c) and 6(d) show a recovery example of a Costas echo passing through the PPS analog link. Similarly, after the recovery, the messy distortions residing in the time-frequency domain plot are clearly eliminated. The large error of the distorted and ground-truth data is decreased, and the MSE averagely decreases by 18.77 dB. The results of LFM echoes passing through the PADC analog link are shown in Fig. 6(e) and 6(f). The distortion effects of the PADC analog link are different from those of the PSS analog link. Beyond the differences in nonlinear amplitude and frequency responses, mismatch distortions are introduced by the multichannelization of the PADC analog link, which does not exist in the case of the PPS. However, the recovery of the RAE is also effective. The decreasing MSE in this case is evaluated to be 20.91 dB. Note that the distortions are clearly eliminated, the MSE improvement is bounded by the random noise and it can be increased with clean waveform inputs. The results validate the general feasibility of the RAE. By merely altering the training datasets, without changing the network structure and hyperparameters, the RAE becomes adaptive to various broadband waveforms and diverse analog links.

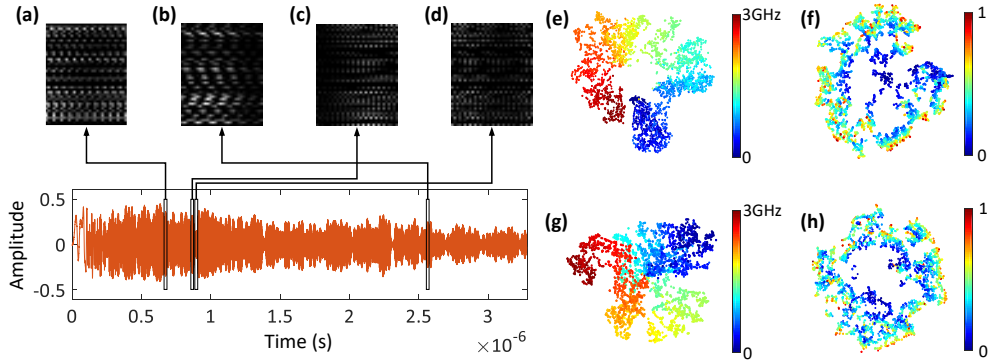

Fig. 7. Feature map analyses with t-SNE. (a-d) Examples of extracted feature maps from different parts of an LFM input waveform. (a)(b) Extracted from low-/high-frequency parts, respectively, and their amplitudes are nearly the same. (c)(d) Extracted from large-/small-amplitude parts, respectively, and their frequencies are nearly the same. (e)(f) t-SNE analyses of frequency dependency and amplitude dependency, respectively. The analyzed RAE is trained with PPS-LFM datasets. The colors represent different frequencies of the input in frequency dependency analysis and different input amplitudes in amplitude dependency analysis. The convergence of similar colors implies that RAE extracts feature maps depending on both frequency and amplitude. (g)(h) t-SNE analyses of frequency dependency and amplitude dependency, respectively, of RAE trained with PADC-LFM datasets. The figure legends are the same as those of (e) and (f).

As shown in Fig. 7, we visualize and investigate the feature maps extracted by the RAE via t-distribution stochastic neighbor embedding (t-SNE) [34]. After several convolutional layers, the input signal is extracted by the RAE to multiple feature maps. Figure 7(a)-7(d) show examples of the extracted feature maps, where different frequencies/amplitudes result in different patterns. As can be seen, high-/low-frequency input does not give corresponding high-/low-frequency feature maps, and large-/small-amplitude input does not correspond to bright/dimmed ones. Even though these feature maps seem incomprehensible from the human perspective, they serve as a tool to analyzing how the RAE gives feature maps depending on different input frequencies and amplitudes. The t-SNE method embeds high-dimensional data to three-dimensional image points for display and understandable analysis. After the dimensionality reduction, the image points are close to each other if their corresponding high-dimensional feature maps are close to each other in Euclidean space. Fig. 7(e) shows the frequency dependency of the RAE that is trained to recover an LFM passing through the PPS analog link. Different colors of the image points represent different frequencies of the input signals. The image points with similar colors are close to each other on the result images, which indicates that the RAE tends to give similar feature maps when the input signals have similar frequencies. The amplitude dependency analysis is shown in Fig. 7(f). The results indicate that the RAE can discriminate different amplitudes and give similar feature maps when the input amplitudes are similar to each other. Figure 7(g) and 7(h) illustrate the frequency dependency and amplitude dependency analyses of the PADC-trained RAE. The results verify that the RAE can learn to automatically extract feature maps depending on input frequencies and amplitudes, which is beneficial for recovering the broadband signals. It is worth noting that most analog links are evaluated via frequency and amplitude responses; the ability to discriminate frequencies and amplitudes strongly suggests that the RAE has considerable potential for more generalized MWP analog links.

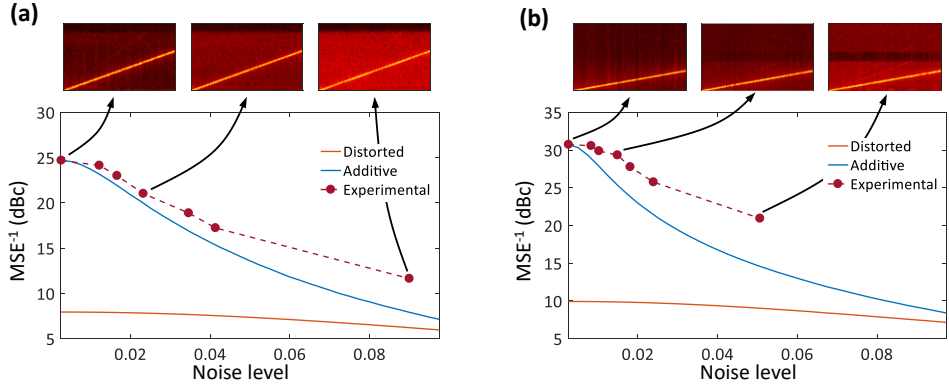

Fig. 8. Noise robustness assessment. Results for (a) PPS system and (b) PADC system. Note that the vertical axis shown in the figure represents the reciprocal of MSE, and the noise level is the mean square root value of the imposed noise. By slowly increasing the AWGN level imposed on the input signal, we draw the MSE results of recovered/unrecovered signals, shown by blue/orange curves. Because distortions mainly contribute to the MSE before recovery, the orange curves drop gradually as the noise level increases. By reducing the averaging times during data acquisition, we get datasets with different levels of experimental noise. The MSE results after recovery are also depicted in the figure with red dots. From left to right, the averaging times are 128, 64, 32, 16, 8, 4, and 1. For every system, three time-frequency-domain plots of test examples are given. They correspond to the recovered signals under 128-/16-/1-time averaging.

As described previously in the data acquisition method, the data used for training and validation is acquired by 128-time averaging. Noise is mitigated significantly so that the datasets are sufficiently clean to demonstrate the ability of distortion recovery. However, in practical applications, data is usually acquired in one step and averaging is inconvenient or impossible. Therefore, we study the performance of the RAE under different noise levels to verify its reliability in practical noisy systems. The noise robustness is assessed in two ways. First, we impose additive white Gaussian noise (AWGN) onto the clean digital-domain input data before it enters the RAE for distortion recovery, and we test the performance of the RAE under different levels of AWGN. Second, by reducing the averaging times of data acquisition, datasets with different actual noise levels are acquired, and we use these actual noisy datasets to test the performance of the RAE. Note that the RAE is not explicitly trained with these noisy datasets or exposed to any noise model, and the RAE for noise robustness assessment is trained only with clean datasets. Results of the noise robustness assessment based on the PPS and PADC datasets are shown in Fig. 8. Note that the errors are induced by both signal distortions and noise. Therefore, in the distorted signals before RAE recovery, distortions mainly contribute to the errors rather than noise. As the noise level increases, the error level does not decrease rapidly. However, after signal recovery, the distortions are eliminated by the RAE and the errors are mainly caused by noise. Consequently, the error level is more influenced by noise level variation. From the results of both the PPS dataset and PADC dataset, the error level increases steadily with the growth of AWGN but it does not surpass the error level before recovery. verifying that the RAE can steadily recover signal distortions in AWGN conditions. Moreover, Fig. 8 shows that, under the same noise level, the RAE recovery performances under actual noise are better than those under the AWGN. Especially, in the assessment of RAE trained with the PADC dataset, the noise-resistive effect is more obvious. Three subplots are provided for each dataset. They show the time-frequency domain plots after signal recovery of the RAE. Their averaging times are $\times 128$, $\times 16$, and $\times 1$, respectively. The increasing noise level provides a thicker noise floor but the RAE still eliminates the distortions effectively without introducing additional artifacts. The results together validate that the RAE is robust against AWGN-noisy data and resistive to actual noisy data. Thus, the effectiveness in practical applications is inferred.

## 5. Conclusion

We have proposed a generally adaptive deep learning scheme that recovers broadband signals in practical defective MWP receiving systems in transceivers. After the generated broadband signals pass through the defective analog links, the actual distortion effects of the analog links are hosted by the distorted signals. A neural network was adopted to learn the end-to-end inverse responses of the distortion effects and the non-distorted time-domain waveforms is recovered. In the experimental demonstrations, we trained the RAE with datasets acquired from various waveforms and diverse analog links. Two MWP systems were set up and two categories of complicated waveforms were chosen to demonstrate the general adaptiveness of the scheme. By altering the training dataset, the RAE could be effectively and adaptively with various waveforms and analog links. The experimental results showed an improvement of ~20 dB in the MSE before and after the recovery. Furthermore, we studied the feature maps extracted by the RAE. The results showed that the trained RAE could construct the inverse frequency and amplitude responses autonomously; thus, it is potentially applicable to more MWP-involved transceivers. For now, the effort to adapt the scheme to another MWP application is to acquire dataset and train the neural network for the new application. However, the development of machine learning technologies will greatly benefit the data acquiring procedure and training procedure. Dataset augmentation technologies like generative adversarial nets [35] and modeled transformation [36, 37] would lessen the burden of data preparation. And advanced training strategies like transfer learning [38, 39] and meta learning [40] would accelerate the training procedure. Furthermore, from the noise robustness assessment results, we concluded that the RAE is steady and robust against noisy inputs and it can perform noise-resistively with actual system noise. Therefore, the proposed scheme is expected to perform effectively and robustly in general MWP systems without additional expert knowledge or auxiliary systems. It enhances the performances of deployed MWP processing systems by only appending a deep learning processor, significantly lowering the cost of further refinement. Consequently, the scheme is expected to facilitate the deployment of next-generation broadband transceiver systems including radars, communications, and microwave imaging.

## Acknowledgments

This work is supported by National Natural Science Foundation of China (grant no. 61822508, 61571292, 61535006).